\documentclass[%
reprint,
superscriptaddress,
 amsmath,amssy
 aps,
prb,
]{revtex4-2}

\usepackage{graphicx}
\usepackage{bm}
\usepackage{ulem}
\usepackage{color}
\usepackage{mathrsfs}
\usepackage[utf8]{inputenc}
\usepackage[T1]{fontenc}
\usepackage{mathptmx}
\usepackage{makecell}
\usepackage{setspace}

\usepackage[colorlinks=true,linkcolor=blue,citecolor=blue]{hyperref}

\begin{document}

\title{Overheated Topological Hall Effect}
	
	\author{Liang Wu}
	\email{liangwu@kust.edu.cn}
	\affiliation{Faculty of Materials Science and Engineering, Kunming University of Science and Technology, Kunming, 650093, Yunnan, China}

	\author{Yujun Zhang}
	\affiliation{Institute of High Energy Physics, Chinese Academy of Sciences, Beijing 100049, China}

\date{\today}

\begin{abstract}
The topological Hall effect (THE) originates from the real-space Berry phase that an electron gains when its spin follows the spatially varying non-trivial magnetization textures, such as skyrmions. Such topologically protected magnetization textures can provide great potential for information storage and processing. Since directly imaging the skyrmions or detecting the magnetic diffraction of skyrmion lattice are significantly more challenging than conducting Hall measurements, THE has been widely used to attest the presence of skyrmions. However, the key feature of THE, namely the bump/dip in the Hall signal is not sufficient proof of THE. Here, we use empirical numerical modeling to demonstrate all possible THE-like signals that two anomalous Hall effect (AHE) signals with opposite signs can superpose. Besides the reproduction of many published results by the numerical model, we propose an exotic {\lq THE\rq} could, in principle, emerge with finely tuned two-channel AHE. The importance of the scrupulous re-examination of the THE observed in experiments cannot be exaggerated. 

\end{abstract}

\maketitle

%\section{\label{sec:level1}First-level heading:\protect\\ The line
%break was forced \lowercase{via} \textbackslash\textbackslash}

The concept of Berry phase has been deeply rooted in modern condensed matter physics, which reveals the role of momentum-space topology in various observable novel effects \cite{Haldane2017, Kosterlitz2017}. The quintessential examples are those notable members in the Hall family, such as the quantum Hall effect and anomalous Hall effect (AHE) \cite{Xiao2010, Nagaosa2010}. Recently, the topology of the non-trivial spin texture in real space would contribute to a non-vanishing Berry phase as well, whose effect on the transverse motion of electrons was dubbed as topological Hall effect (THE) \cite{Bruno2004}. The widely-used criterion of the presence of THE is the additional bumps/dips along with the AHE. However, it has been demonstrated in many papers, say Refs. \cite{Kan2018, Gerber2018, Wu2020}, that the above criterion is no longer sufficient when there exist multiple conduction channels with opposite AHE signs. Although a large number of publications have discussed this issue to date, there are still far too many papers reporting THEs without considering the possible artifacts. 

Here, simple numerical modeling was utilized to extend and generalize the two-channel AHE situations, which not only reproduces the published {\lq THE\rq} results, but also puts forward the {\lq last\rq}  possible artificial THE signals not experimentally observed yet. We underline that THE-like Hall signals should be taken more seriously when claiming the existence of THE and related skyrmions.

\bigskip

THE-like Hall signal can be mimicked by the superposition of two AHE signals with opposite signs. Similar to the magnetization versus field hysteresis, an empirical model of
$R^\mathrm{I}_{\mathrm{AHE}}=M_\mathrm{I}\mathrm{tanh}(\omega_\mathrm{I}(H-H^\mathrm{I}_\mathrm{c}))$, can be used to model the AHE signals, where $M_\mathrm{I}$, $\omega_\mathrm{I}$ and $H^\mathrm{I}_\mathrm{c}$ stand for the AHE magnitude, slope parameter at the coercive field, and coercive field. An additional AHE channel can be represented as
$R^\mathrm{II}_{\mathrm{AHE}}=M_\mathrm{II}\mathrm{tanh}(\omega_\mathrm{II}(H-H^\mathrm{II}_\mathrm{c}))$. Thus the overall AHE can be presented by the following equation,

\begin{equation*}
\begin{aligned}
R^{\mathrm{tot}}_{\mathrm{AHE}}= & M_\mathrm{I}R^\mathrm{I}_{\mathrm{AHE}}\mathrm{tanh}(\omega_\mathrm{I}(H-H^\mathrm{I}_\mathrm{c})) \\
& + M_\mathrm{II}R^{\mathrm{II}}_{\mathrm{AHE}}\mathrm{tanh}(\omega_{\mathrm{II}}(H-H^{\mathrm{II}}_\mathrm{c})) 
\end{aligned}
\end{equation*}

\begin{figure*}
	\centering
	\includegraphics[width=\linewidth]{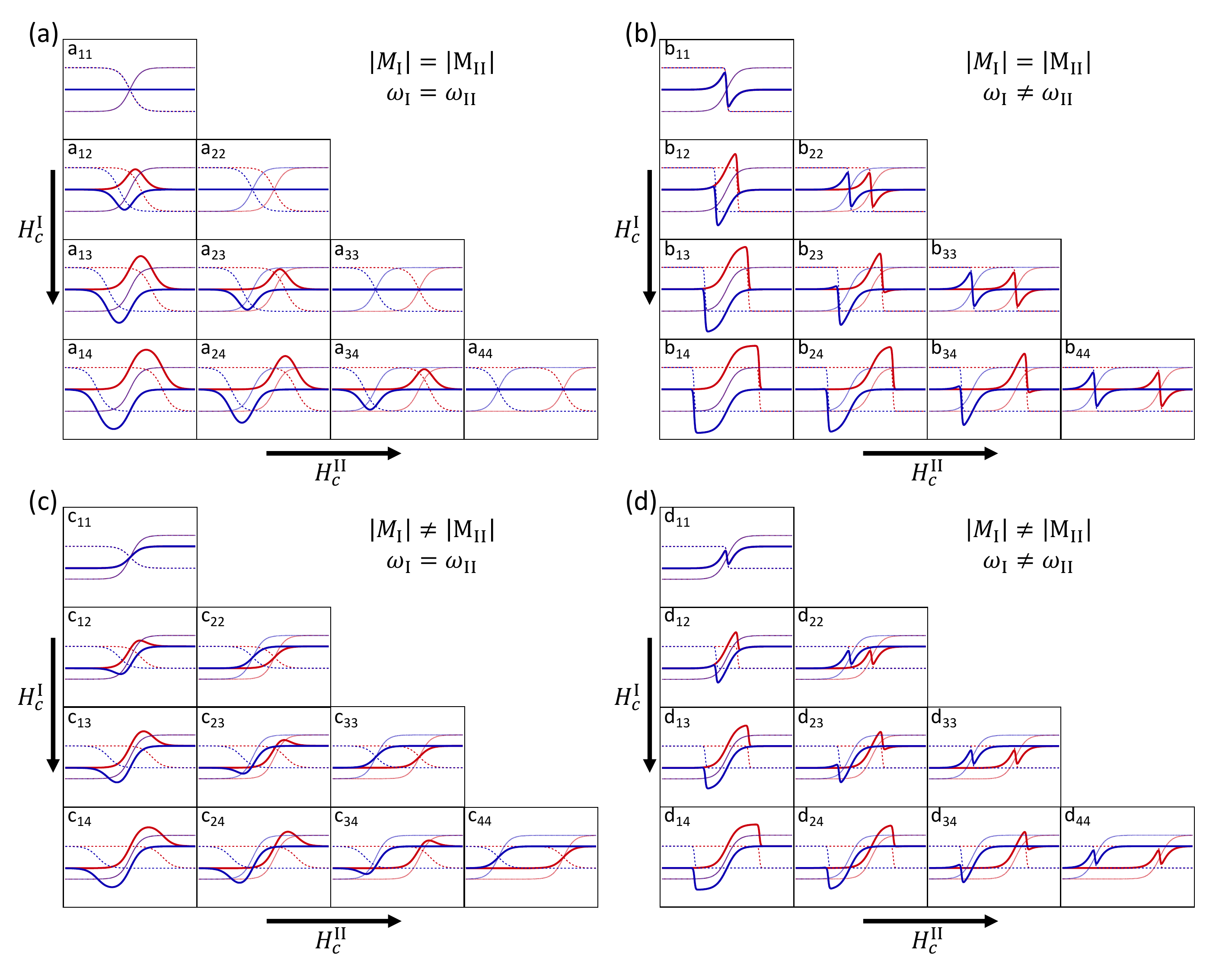}
	\caption{\textbf{Numerical modeling of the superposition of two opposite AHE signals.} The doted, dashed and solid lines represent the positive, negative and total Hall channels, respectively. The colors indicate the directions of field sweep (red: increase, blue: decrease).  (a) and (b) are situations with $\omega_\mathrm{II}>\omega_\mathrm{I}$ and $\omega_\mathrm{II}<\omega_\mathrm{I}$, respectively, $M_\mathrm{I}$ and $M_\mathrm{II}$ are fixed at 5 and $-3$. (c) special cases with $H^\mathrm{I}_\mathrm{c}=H^\mathrm{II}_\mathrm{c}=0$ for the top two panels, while $M_\mathrm{I}=M_\mathrm{II}$ for the bottom panel. Note that the behaviors of the overall AHE signal will be reversed about the $y$-axis when $|M_\mathrm{I}|<|M_\mathrm{II}|$.}
	\label{Fig1}
\end{figure*}

To simplify the discussion, we set $M_\mathrm{I}>0$ and $M_\mathrm{II}<0$ to ensure the two channels are of opposite signs. Thus, the characteristic shape of the overall AHE would only be affected by the relative magnitude of these parameters. By exhaustive enumeration, we systematically studied all four cases with (a) $|M_\mathrm{I}|=|M_\mathrm{II}|$ and $\omega_\mathrm{I}=\omega_\mathrm{II}$, (b) $|M_\mathrm{I}|=|M_\mathrm{II}|$ and $\omega_\mathrm{I}\ne\omega_\mathrm{II}$, (c) $|M_\mathrm{I}|\ne|M_\mathrm{II}|$ and $\omega_\mathrm{I}=\omega_\mathrm{II}$, (d) $|M_\mathrm{I}|\ne|M_\mathrm{II}|$ and $\omega_\mathrm{I}\ne\omega_\mathrm{II}$, by varying the relative magnitude of their coercive fields $H_\mathrm{c}$, as shown in Fig. \ref{Fig1}.  Without loss of generality, here we set $|M_\mathrm{I}|>|M_\mathrm{II}|$ for $|M_\mathrm{I}|\ne|M_\mathrm{II}|$, which indicates the magnitude of AHE for channel I is larger than that of channel II.  While we set  $\omega_\mathrm{I}<\omega_\mathrm{II}$ for $\omega_\mathrm{I}\ne\omega_\mathrm{II}$,  which means channel II has a steeper transition than channel I near $H_\mathrm{c}$. 

As immediately seen in Fig. \ref{Fig1}, the THE-like bumps/dips can generally be triggered in many cases. Hall signals similar to subfigures of $ \mathrm{a}_{34} $, $ \mathrm{b}_{24}$,  $ \mathrm{c}_{23}$, $ \mathrm{c}_{24}$, and  $ \mathrm{d}_{11} $ are the most widely observed ones in experiments, see the review article of Ref. \cite{Kimbell2022}. To date, only {\lq THE\rq} similar to subfigure $\mathrm{b}_{44}$  has not been experimentally observed yet. 

The AHE is both sensitive to the intrinsic electronic structure near the Fermi level and extrinsic sample inhomogeneity, the coexistence of two opposite AHE channels could occur in many materials and heterostructures. As for SrRuO$_3$, its Berry curvature is so sensitive to many tuning knobs, that the AHE sign could readily be reversed by thickness \cite{Wu2020}, temperature\cite{Fang2003}, disorder level \cite{Wu2020}, doping \cite{Fang2003, Li2020}, strain \cite{Tian2021}, interfacial effects \cite{Matsuno2016, Wang2018}, and electric gating \cite{Ohuchi2018}. Therefore, a small amount of inhomogeneity can blend positive and negative AHE signals to produce an artificial THE near the sign reversal regions. For instance, Wu and his colleagues firstly noticed that SrRuO$_3$ shows a sign reversal at a critical thickness of 4 unit cell (uc), even one uc thickness fluctuation can mix the positive AHE from 4 uc and negative AHE from 5 uc (or thicker regions) to induce the bump in Hall signal in a nominal 4 or 5 uc sample \cite{Wu2020}. This easily-tuned sign of AHE might account for the ubiquitous  {\lq THE\rq} in SrRuO$_3$. Most recently, such issues becomes more seriously taken, Tai et. al. proposed protocols to distinguish the two-component AHE from the THE \cite{Tai2022}.

\bigskip

As a newly prominent research field, dedicated efforts have been devoted to the study of THE. However, the authenticity of THE has long been overlooked. We accentuate that different forms of THE can be imitated with tricky two-channel opposite AHE, which is a natural result in systems with AHE sign reversal.

\bigskip
L.W. acknowledges support from Natural Science Foundation of
China (Grant No. 52102131).

\bibliography{ref}
\end{document}